\documentclass[12pt]{iopart}
\usepackage{epsfig,graphicx,tabularx}
\usepackage{amssymb}
\usepackage{psfrag}
\usepackage{graphicx}
\usepackage{graphics}
\usepackage{epsfig}
\usepackage{epstopdf}
\usepackage{hyperref}
\usepackage{bm}
\usepackage{verbatim,color,ulem}
\usepackage{color}

\newcommand{\beq}{\begin{equation}}
\newcommand{\eeq}{\end{equation}}
\newcommand{\beqa}{\begin{eqnarray}}
\newcommand{\eeqa}{\end{eqnarray}}
\newcommand{\rb}{$^{87}$Rb}
\newcommand{\kq}{$^{41}$K}
\definecolor{drkgreen}{rgb}{0.0, 0.5, 0.0}

\newcommand{\gr}[1]{\textcolor{drkgreen}{#1}}

\begin{document}

\title{Collision of impurities with Bose-Einstein condensates}

\author{F. Lingua$^{1,2}$, L. Lepori$^{3,4}$, F. Minardi$^{5,6}$,
V. Penna$^{1}$, and L. Salasnich$^{5,7}$}
\address{$^{1}$Dipartimento di Scienza Applicata e Tecnologia,
Politecnico di Torino, Corso Duca degli Abruzzi 24, I-10129 Torino, Italy \\
$^{2}$Department of Physics, Clark University, Worcester, Massachusetts
01610, USA\\
$^{3}$Dipartimento di Scienze Fisiche e Chimiche, Universit\'a
dell'Aquila, via Vetoio, 67010 Coppito, Italy\\
$^{4}$INFN, Laboratori Nazionali del Gran Sasso, Via Acitelli 22,
67100 Assergi, Italy \\
$^{5}$Istituto Nazionale di Ottica, INO-CNR, via Giovanni Sansone 1, 50019 Sesto Fiorentino, Italy\\
$^{6}$LENS European Laboratory for Non-Linear Spectroscopy, and Dipartimento di Fisica e Astronomia,
via Nello Carrara 1, 50019 Sesto Fiorentino, Italy \\
$^{7}$Dipartimento di Fisica e Astronomia ``Galileo Galilei'' and CNISM,
Universit\`a di Padova, Via Marzolo 8, 35131 Padova, Italy}

\date{\today}

\begin{abstract}
Quantum dynamics of impurities in a bath of bosons is a
long-standing problem of solid-state, plasma, and atomic physics.
Recent experimental and theoretical investigations with ultracold atoms
focused on this problem, studying atomic impurities immersed
in a atomic Bose-Einstein condensate (BEC) and for various relative
coupling strengths tuned by the Fano-Feshbach resonance technique.
Here we report extensive numerical simulations on a closely related problem:
the collision between a bosonic impurity made of few \kq\ atoms
and a BEC made of \rb\ atoms in a quasi one-dimensional configuration
and under a weak harmonic axial confinement.
For small values of the interspecies interaction strength
(no matter the sign of it), we find that the impurity, which
starts from outside the BEC, simply oscillates back and forth the BEC cloud,
but the frequency of oscillation depends
on the interaction strength. For intermediate couplings,
after a few cycles of oscillation the impurity is captured by the BEC
and strongly changes its amplitude of oscillation.
In the strong interaction regime, if the interspecies interaction is attractive, a local maximum (bright soliton) in the density of BEC occurs
where the impurity is trapped; instead, if the interspecies interaction is repulsive, the impurity is not able to enter in the BEC cloud
and the reflection coefficient is close to one. On the other hand, if the initial displacement of the impurity is increased, the impurity is able to penetrate in the cloud leading to the appearance of a moving hole (dark soliton) in the BEC.
\end{abstract}

\maketitle

\section{Introduction}

In 1933 Landau introduced the concept of polaron, an
electron whose effective mass is affected by the coupling
with the quantized lattice vibrations (phonons)
of a crystal \cite{landau}. Later, Fr\"olich derived a field-theoretical
polaron Hamiltonian that covers all coupling strengths between
the electron and the phonons \cite{frolich}. The basic properties
of polarons are now established (see, for instance,
the review \cite{review-polaron})
but the interest in the polaron dynamics and, more generally,
in the dynamics of impurities interacting with a bosonic bath
has recently gone through a vigorous revival, mainly in the context
of ultracold atomic gases.
In Ref. \cite{catani2012}, localized bosonic impurities, made
of few \kq\ atoms, have been created
in a one-dimensional (1D) configuration
and their interactions with a Bose-Einstein condenstate (BEC) of \rb\
atoms have been investigated by using a Fano-Feshbach resonance
to tune the impurity-boson scattering length.
More recently tunable BEC impurities, i.e. atomic impurity in a
cloud of ultracold Bose-Einstein condensed atoms, have been obtained by other
two experimental groups \cite{jin,arlt}.
The Bose polaron problem has been addressed theoretically
by using different techniques: quantum Langevin equation \cite{catani2012},
mean-field theory with coupled Gross-Pitaevskii
equations \cite{jaksch2012,pelster2016}, time-dependent variational mean-field
for lattice polarons \cite{grusdt,demlerLattPol}, Feynman path integral and
Jensen-Feynman variational principle \cite{tempere},
T-matrix \cite{rath2013} and perturbation \cite{christensen2015} approaches,
variational wavefunction \cite{das2014,levinsen2015}, and quantum
Monte Carlo schemes \cite{ardila2015}. Recently, the transition of the impurity from the polaron to the  soliton state has been studied in \cite{poltosol2015} by combining the Fr\"olich Hamiltonian picture with the Landau-Brazovskii theory for first-order phase transitions.
All these theories work quite well in the weak-coupling regime but
show some deviations with respect to
recent experiments \cite{jin,arlt} in the strong-coupling regime.
However, a nonperturbative renormalization-group
approach \cite{demler} seems able to give a reliable and
unified picture of the Bose polaron problem from
weak to strong coupling, also in one-dimensional
configurations \cite{giorgini2017}.

A common feature of the above investigations is that the
impurity always remains inside the bosonic bath. In this paper
we study a different but closely related non-equilibrium problem:
the collision between a \kq\ impurity and a \rb\ BEC, where
initially the impurity is outside the bosonic bath.
In Section II we introduce the physical system, that is a quasi
one-dimensional BEC of 300 \rb\ atoms located at the minimum of
a weak harmonic axial trap and an impurity of 5  atoms that
starts from the edge of the BEC cloud. In Section III we define
the two coupled one-dimensional time-dependent Gross-Pitaevskii
equations which are used to perform the numerical simulations.
In Section IV we discuss our theoretical predictions
which display a very rich phenomenology crucially
depending on the impurity-BEC strength.
Indeed our numerical simulations unequivocally show that the collision
dynamics gives rise to a variety of highly-nonlinear effects
such as dark and bright solitons, impurity trapping, and coupled
oscillations in the confining trap, which,
due to their macroscopic character, pave the way to
forthcoming experiments. The paper is concluded by Section V.

\section{Properties of the system}

We consider a setup very close to the one
realized experimentally in Ref. \cite{catani2012}:
a bosonic cloud of \rb\ atoms and a bosonic impurity
made of \kq\ atoms in a 1D harmonic confinement. This quasi-1D configuration is
obtained by a weak optical confinement along one direction and
a strong optical confinement along the other two transverse
directions. The atoms interact by intra-species and inter-species interactions.
In particular, in this paper the inter-species
interaction is controlled by the magnetic Feshbach resonance. On the contrary,
the intra-species interactions are always repulsive and close to their
background values in the range of magnetic field used to exploit the
Feshbach resonance. In a realistic experiment,
the variation of the inter-species s-wave scattering length $a_{Rb,K}$
between \rb\ and \kq\ atoms by a Feshbach
resonance can induce also variations of intra-species scattering lengths
$a_{Rb}$ and $a_K$. However, these variations are quantitatively
negligible \cite{catani2012}. In our simulations we use the following values
for the 3D scattering lengths:
$a_{Rb} = 100 \, a_0$ \cite{egorov2013} and
$a_K = 63 \, a_0$ \cite{falke2008}
with $a_0$ the Bohr radius.
We work with $N_{Rb}=300$ atoms for Bose-Einstein condensate
and $N_K=5$ atoms for the impurity. Unlike the experiment of
Ref. \cite{catani2012}, in this work we assume zero temperature throughout.
For \rb\ atoms the frequencies of transverse and axial harmonic confinement are
$\omega_{\perp Rb}= 2\pi \times 34\times10^{3}\, \mathrm{Hz}$
and $\omega_{\parallel Rb}=2\pi \times 62\, \mathrm{Hz}$.
For \kq\ atoms they are instead
$\omega_{\perp K}=2\pi \times 45\times10^{3}\, \mathrm{Hz}$
and $\omega_{\parallel K}=2\pi \times 87\,\mathrm{Hz}$.

Two lengths characteristic of the considered problem can be defined
naturally for each species:
the longitudinal and transverse harmonic oscillator lengths, respectively
$a_{||,s} = \sqrt{{\hbar}/({m_s \, \omega_{|| \, s}})}$ and
$a_{\perp, s}=\sqrt{{\hbar}/({m_s \, \omega_{\perp , s}})}$ with $s=$Rb, K.
Due to the relations $\omega_{\perp Rb} \gg \omega_{\parallel Rb}$
and $\omega_{\perp K} \gg \omega_{\parallel K}$ between the confinement
frequencies, the system behaves effectively as one-dimensional.
The one-dimensional scattering lengths characterizing
the intra-species interactions
are obtained from the three-dimensional ones
by the Olshanii formula \cite{olshanii}:
$a_{s,1D} = - (a_{\perp}^2/a_{s})(1- C (a_{s}/a_{\perp, s}))$
with $C \simeq 1.4603/\sqrt{2}$. In this way, the 1D intra-species
interaction strengths are given by
\beqa
g_{1} = - \frac{2 \, \hbar^2}{m_{Rb} \, a_{Rb,1D}}
\\
g_{2} = - \frac{2 \, \hbar^2}{m_{K} \, a_{K,1D}}
\eeqa
with $g_1 = 2.365$ J$\cdot$m and $g_2= 0.8598\, g_1$;
while the 1D inter-species interaction strength between \rb\ and
\kq\ atoms, $g_{12}$, is numerically calculated following
the analysis laid out in \cite{peano2005}.
Following Ref. \cite{thalhammer2009}, we determine the 3D
inter-atomic scattering
length $a_{Rb,K}$ as a function of the magnetic field $B$ near the
Feshbach resonance at $78.2$ Gauss. We then derive
the corresponding 1D inter-atomic
strength $g_{12}$.

\begin{figure}
\centering
\includegraphics[width=0.7\textwidth]{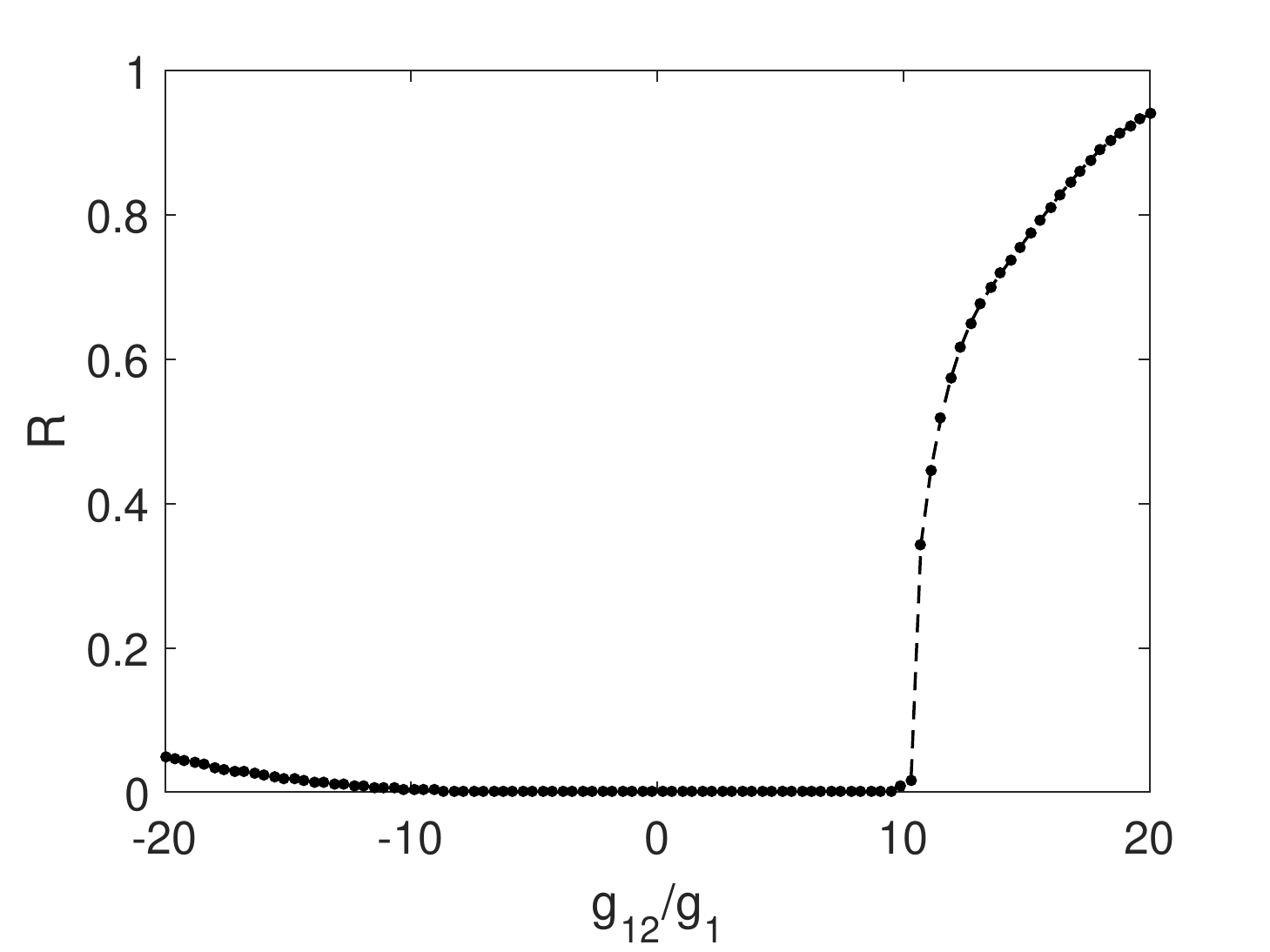}
\caption{Reflection coefficient $R$ of $^{41}$K impurity on
$^{87}$Rb cloud as a function of the adimensional
inter-atomic strength $g_{12}/g_1$ at time $t=\pi/\omega_{||K}$.}
\label{f1}
\end{figure}

\section{Theoretical approach}

We describe the system under investigation in the frame of mean-field
theory for both components.  In the configuration of
Ref.\cite{catani2012}, the transverse confinement corresponds to
a harmonic-oscillator length of approximately
$1150 a_0$ for the majority component
(Rb). As a consequence, the 1D gas parameter, i.e. the ratio of the healing
length to interparticle distance, is $\xi/d =
a_{\perp, Rb}/\sqrt{8 a_{Rb} d} \lesssim 1$,
justifying the use of mean-field theory for the $^{87}$Rb cloud.
In fact, $^{87}$Rb atoms are in the 1D weak-coupling quasi-condensate
regime, where the exact Lieb-Liniger theory of 1D bosons with repulsive
contact interaction reduces to the 1D Gross-Pitaevskii model \cite{book-stringari,Salasnich2016}.
We complement the complex wavefunctions $\psi_{Rb}(x,t)$
with a similar wavefunction $\psi_{K}(x,t)$  for the impurity and impose
that they satisfy the coupled Gross-Pitaevskii equations (GPEs)
\beqa
i\hbar\frac{\partial \Psi_{Rb}}{\partial t}&=& \left [
-\frac{\hbar^2}{2m_{Rb}}{\partial^2\over \partial x^2}
+ \frac{1}{2}m_{Rb} \, \omega_{||Rb}^2 \, x^2
+  g_{1}|\Psi_{Rb}|^2  + g_{12} |\Psi_{K}|^2 \right ]\Psi_{Rb}
\label{GPE1}
\\
i\hbar\frac{\partial \Psi_{K}}{\partial t}
&=& \left [
-\frac{\hbar^2}{2m_{K}}{\partial^2\over \partial x^2}
+ \frac{1}{2}m_{K} \, \omega_{||K}^2 \, x^2 +  g_{2}|\Psi_{K}|^2  +
g_{12} |\Psi_{Rb}|^2 \right ]\Psi_{K}  \, ,
\label{GPE2}
\eeqa
with
\beqa
N_{Rb} = \int |\Psi_{Rb}(x,t)|^2 \ dx \label{NR}
\\
N_{K} = \int |\Psi_{K}(x,t)|^2 \ dx \;\label{NK}
\eeqa
the atom numbers, 
which remain constant during the time evolution. In our model also
the $^{41}$K impurity is described by a 1D GPE, but the very small number
of $^{41}$K atoms implies that the nonlinear term proportional to $g_2$
is extremely small. We have verified that setting $g_2=0$ the numerical
results of Section 4 are practically the same.
%
%
The key parameter is
instead $g_{12}$, that is the strength of the density-density coupling
between $^{87}$Rb cloud and $^{41}$K impurity.
An accurate estimate of the axial density profile of the $^{87}$Rb cloud
(at the initial time and with good approximation also during the dynamics)
can be achieved by the local density approximation, leading to
\beq
\rho_{Rb}(x,0) = |\Psi_{Rb}(x,0)|^2 \simeq
\left\{ \begin{array}{ll}
{1\over g_1} \left(
\mu_{Rb} - \frac{1}{2} \, m_{Rb} \, \omega_{\parallel Rb}^2 \, x^2
\right)
& \mbox{ if $|x| \leq R_{Rb}$ } \\
        0 & \mbox{ elsewhere }
\end{array} \right.  \, ,
\eeq
where
\begin{equation}
R_{Rb} = \sqrt{({2 \, \mu_{Rb}})/({m_{Rb} \, \omega_{\parallel Rb}^2})}
\end{equation}
\begin{equation}
\mu_{Rb}=((3 g_{Rb} N_{Rb} \sqrt{m_{Rb}
\omega_{\parallel Rb}^2})/(4\sqrt{2}))^{2/3}
\end{equation}
are the Thomas-Fermi radius of the Rb cloud
and the corresponding chemical potential respectively. Inserting the
numerical values for the various parameters one finds
$\mu_{Rb} = 75 \hbar \omega_{\parallel Rb}$ 
and $R_{Rb} = 12.2 a_{\parallel Rb}$, 
where $a_{\parallel Rb}=\sqrt{\hbar / m_{Rb}\omega_{\parallel Rb}}=1.37 \mu$m.

\section{Numerical results}

For the \kq\ impurity at the initial time of considered evolution, we assume
for most of the considered cases a Gaussian wavefunction
\beq
\Psi_K(x,0) = {N_{k}\over \pi^{1/4} \sigma^{1/2}} \ e^{-(x+d_0)^2/(2 \sigma^2)}
\eeq
centered at a distance $d_0 = 14.6 a_{\parallel Rb} =20 \mu$m
from the origin of the axial harmonic trap.
The width $\sigma$ is chosen
to be equal to $a_{\parallel Rb}$. 
Similarly, in the following (simulations included)
all the lengths will be measured in unit of $a_{\parallel Rb}$.

Eqs. (\ref{GPE1}) and (\ref{GPE2}) are solved numerically
by using a finite-difference predictor-corrector Crank-Nicolson
algorithm \cite{salas1998}.
The center of mass of the \rb\ BEC remains practically constant
while the center of mass of \kq\ impurity, which is initially
outside the \rb\ BEC, evolves in time due to the axial harmonic
potential and its dynamics strongly depends on interaction strength $g_{12}$
between the impurity and the $^{87}$Rb condensate.
As expected if $g_{12}=0$ the \kq\ impurity crosses the
\rb\ cloud without perturbing it (see movie \cite{movie-0} for details).
In this case the \kq\ impurity
simply oscillates back and forth with oscillation frequency $\Omega$,
that is exactly the frequency $\omega_{\parallel K}$ of the axial harmonic
confinement for \kq\ atoms.

In Fig. \ref{f1} we plot the reflection coefficient $R$
of the \kq\ impurity for different values of the inter-atomic strength
$g_{12}/g_1$ at time $t=\pi/\omega_{||K}$.
The reflection coefficient is computed as
\begin{equation}
R(t)=\frac{1}{N_{K}}\int_{-\infty}^{0}|\Psi_{K}(x,t)|^2 \ dx
\end{equation}
where $N_{K}$ is the number of \kq\ atoms. The
coefficient $R$ equals the unity when the \kq\ impurity is entirely
confined in the left side of the space domain ($x<0$), while $R$ vanishes when
the impurity is completely in the right side of the domain ($x>0$).
As expected, at the time $t=\pi/\omega_{|| K}$, which corresponds
to half period of oscillation, in the absence of inter-atomic
interaction ($g_{12}=0$) the reflection coefficient is zero.
For strongly repulsive ($g_{12}/g_1 \gg 1$) interactions close to the
resonance (i.e. for $B \simeq 78.2$ Gauss),
Fig. \ref{f1} shows instead that the impurity is completely confined out
the cloud (reflected back) leading to reflection
coefficient $R \simeq 1$. For weakly attractive $g_{12}/g_1$, the coefficient
$R$ always lies at very small values close to zero. However, for strongly
attracting values of $g_{12}/g_1$ it is possible to notice a small increase
in $R$. This 
occurs because, due to the strongly-attractive
interaction, the \kq\ wave-packet tends to be captured 
toward the center of mass
of the \rb\ cloud and, at the time $t=\pi/\omega_{|| K}$, a fraction of it is still in the left side $(x<0)$ of the space domain. For the same reason,
the oscillation frequency of the strongly-attractive case
reaches higher values with respect to $\omega_{\parallel K}$, as shown
in the first panel of Fig. \ref{f2}.

\subsection{Weak-coupling and periodic motion of impurity}

\begin{figure}
\centering
\includegraphics[width=0.7\textwidth]{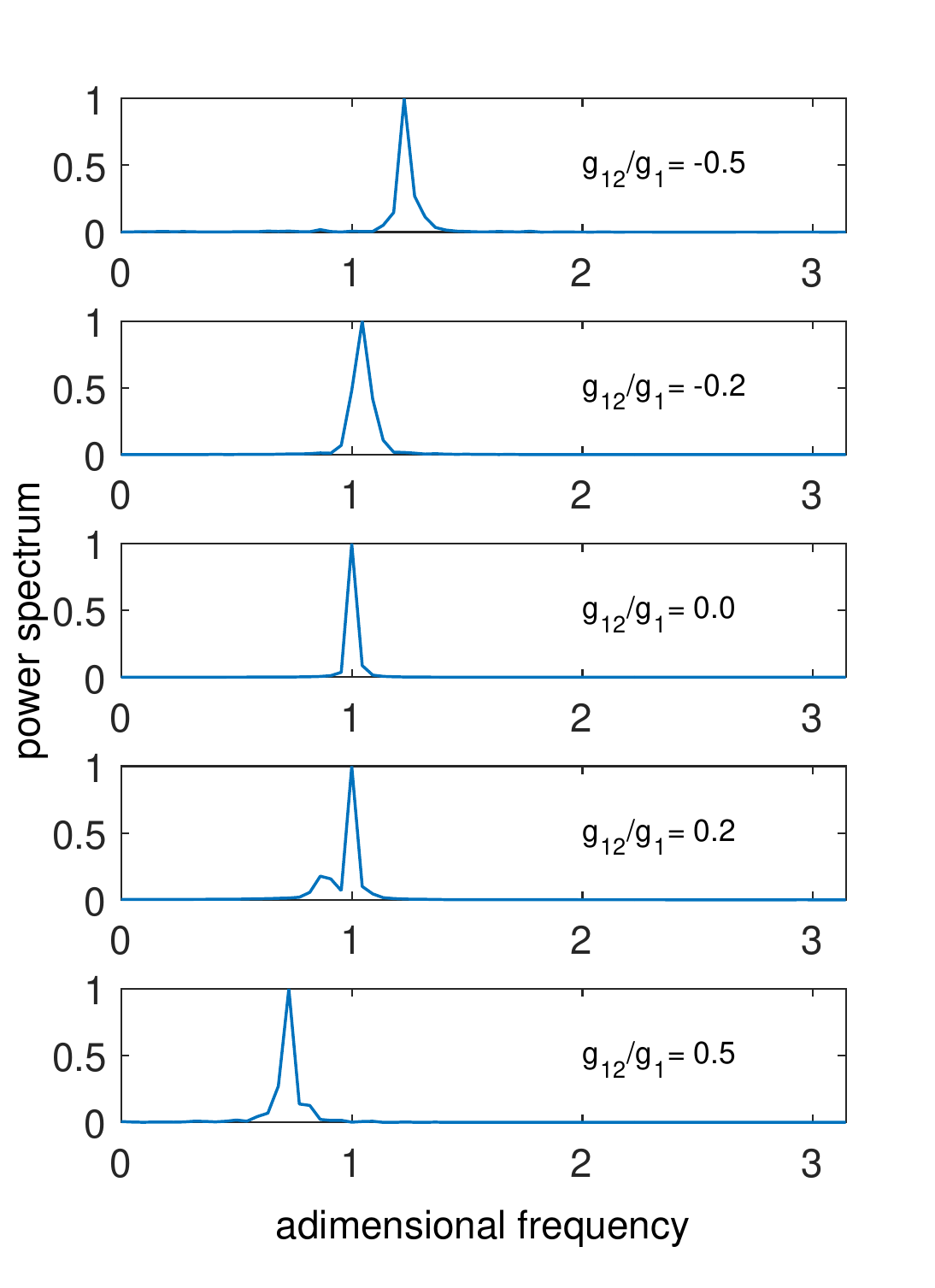}
\caption{Power spectrum $|{\tilde x}(\omega)|^2$ vs adimensional
frequency $\omega/\omega_{|| K}$ of the Fourier
transform ${\tilde x}_{cm}(\omega)$ of the center-of-mass position
$x_{cm}(t)$ of the \kq\ impurity. Results obtained for different
values of the adimensional interatomic strength $g_{12}/g_1$
between $^{41}$K and $^{87}$Rb atoms. In each panel $|{\tilde x}(\omega)|^2$
is scaled so that its absolute maximum is equal to one.
$\omega_{|| K}$ is the frequency of the
axial harmonic confinement for $^{41}$K atoms.}
\label{f2}
\end{figure}

For small values of the inter-atomic strength, i.e. for $|g_{12}|/g_1
\lesssim 1$ the $^{41}$K the impurity simply oscillates
back and forth inside the $^{87}$Rb cloud: However,
the frequency $\Omega$ of oscillation depends on $g_{12}$.

To determine $\Omega$ we calculate
the center-of-mass position $x_{cm}(t)$
of the $^{41}$K impurity as a function of time $t$.
Then we perform the Fourier transform
\beq
{\tilde x}_{cm}(\omega) = \int x_{cm}(t) \ e^{i \omega t} \ dt \;
\eeq
of $x_{cm}(t)$ and plot its power spectrum $|{\tilde x}_{cm}(\omega)|^2$
vs $\omega/\omega_{|| K}$ in Fig. \ref{f2}.
The panels of this figure are obtained
for different values of the inter-atomic strength $g_{12}$ between
\kq\ and \rb\ atoms. The figure clearly shows that,
as expected, for $g_{12}=0$ (middle panel)
there is only one peak centered at $\omega/\omega_{|| K} = 1$
and, consequently, the center of mass oscillates
at the frequency $\Omega= \omega_{|| K}$, that is the frequency
of axial harmonic confinement of the $^{k1}$K atoms.

Upper panels of Fig. \ref{f2} reveal
that for small and negative (attractive) values of $g_{12}/g_1$
the frequency $\Omega$ of oscillation
of the $^{41}$K impurity increases.
Instead, for small and positive (repulsive) values of
$g_{12}/g_1$ (lower panels) a second mode appears at a lower frequency. This
second mode becomes dominant growing $g_{12}/g_1$.

\subsection{Intermediate coupling and impurity trapping}

\begin{figure}
\centering
\includegraphics[width=0.9\textwidth]{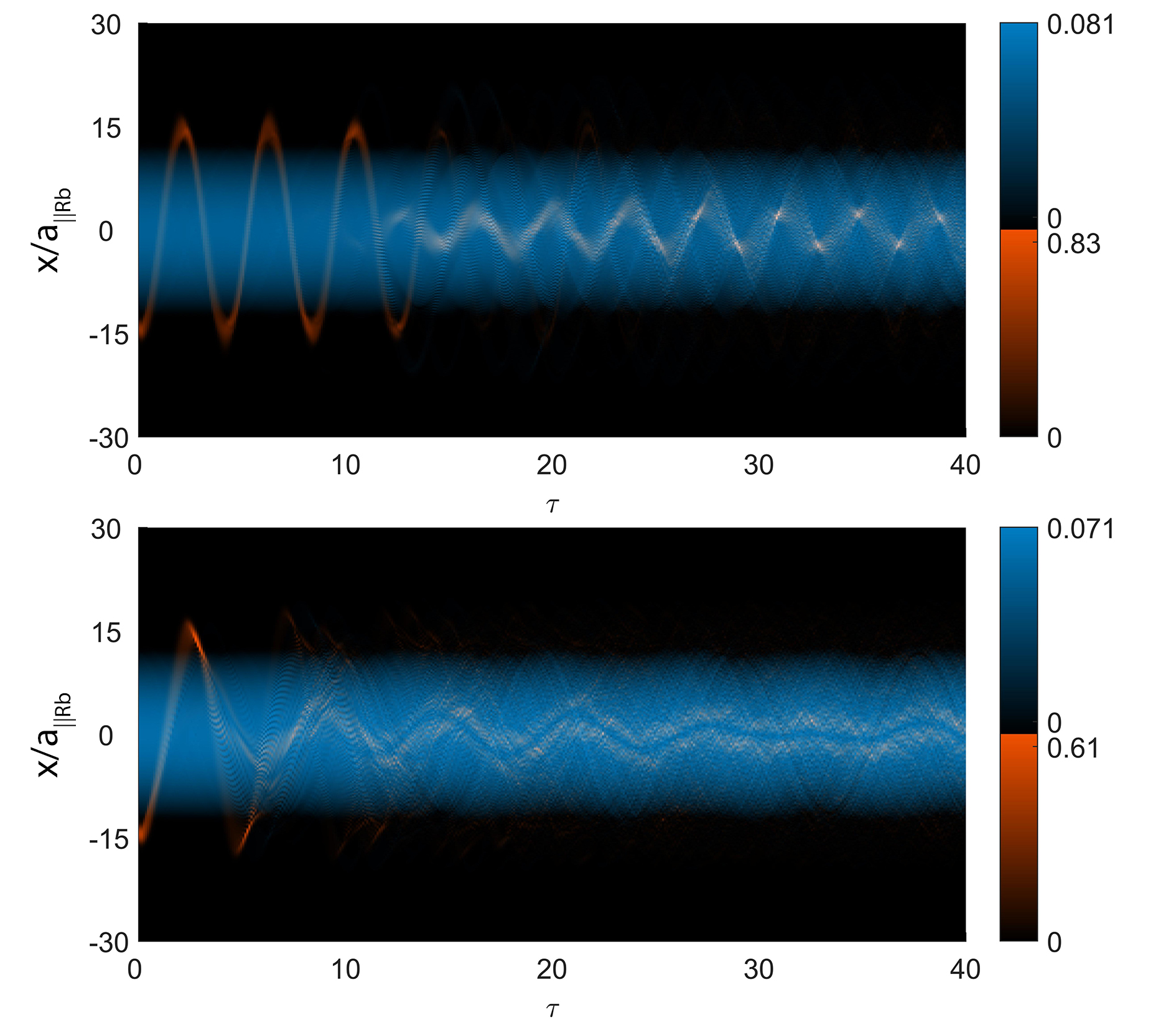}
\caption{Upper panel: normalized distribution for $^{87}Rb$,
$|\Psi_1(x,\tau)|^2/N_{Rb}$ (light-blue) and for $^{41}K$,
$|\Psi_2(x,\tau)|^2/N_K$ (orange), as a function of the adimensional
space $x/a_{\parallel Rb}$ and adimensional time $\tau = \omega_{\parallel Rb} t$,
for  $g_{12}=-0.5$. Lower panel: same quantities for $g_{12}=0.5$.
Here $a_{\parallel Rb}=1.37$ $\mu$m and $\omega_{\parallel Rb} = 389.6$ Hz.}
\label{f3}
\end{figure}

For intermediate couplings, i.e. around $|g_{12}|/g_1\simeq 0.5$
no matter the sign of $g_{12}$, after a few cycles of oscillation
the $^{41}$K impurity is captured by the $^{87}$Rb cloud
and it strongly changes its amplitude of oscillation.
This phenomenon is shown in Fig. \ref{f3}, where the contour plot
of $^{41}$K and $^{87}$Rb density profiles is reported.

A difference between the attractive and the repulsive case is visible
in the shape of the density profile of the impurity at fixed time.
Indeed the repulsive case (lower panel) shows a mode of oscillation of the
impurity featuring a density profile characterized
by two main lobes for the distribution, with always an empty-region at
the center of the oscillating wave packet. In other words, the impurity
profile  features, along the $x$-direction,  two main maxima and one
central minimum. On the other hand, the attractive case (upper panel)
 displays an oscillation mode in which the wave
packet features only a single lobe: a density profile featured by
only one maximum.
  Moreover, in spite of the relevant magnitude of the attraction, no significant quantum
  reflection phenomena occurring for rapidly varying attractive potentials
  \cite{pasquini2004,Cornish_Qrefl} are 
  observed at the boundaries of the BEC condensate,
  neither the consequent vortex formation
  \cite{sheard2005}. 
  The latter absence, also holding for abrupt
  repulsive potential \cite{gardiner2006}, is ascribable to
  the one-dimensionality of our simulations, 
  as well as to the small size of the
  impurity.

The full dynamics of the attractive and repulsive cases is well
illustrated in Fig. \ref{f3} and it can also be seen in
the movies \cite{movie-weakAttr} and \cite{movie-weakRep}, respectively.

\subsection{Strong-coupling and solitary waves}

\subsubsection{Repulsive inter-atomic strength.}

\begin{figure}
\centering
\includegraphics[width=0.9\textwidth]{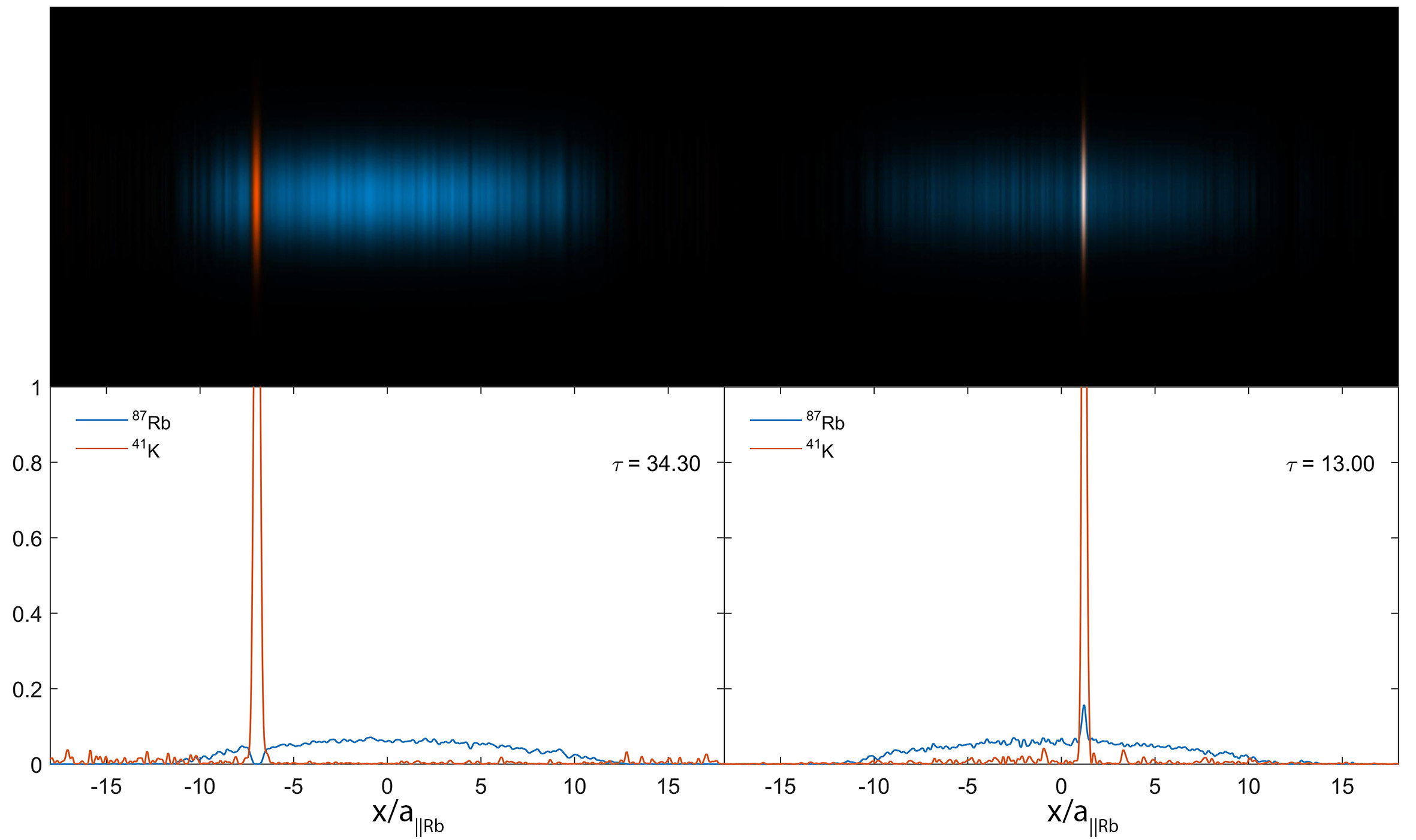}
\caption{Left panels: dark soliton in the \rb\ cloud
induced by the $^{41}$K impurity, for $g_{12}/g_1 = 4$.
Right panels: \rb\ bright soliton
induced by the \kq\ impurity, for $g_{12}/g_1 = -2$.
Upper panels give the contour plot of the normalized
density profiles in the $(x,y)$ plane. Lower panels give
the corresponding normalized axial density profiles
as a function of the adimensional axial coordinate $x/a_{\parallel Rb}$.
$\tau= \omega_{\parallel Rb} t$ is the adimensional time
with  $\omega_{\parallel Rb} = 389.6$ Hz and $a_{\parallel Rb}=1.37$ $\mu$m. Frames are taken from the movies \cite{movie-dark} and \cite{movie-bright}. }
\label{f4}
\end{figure}

When the inter-atomic interaction is repulsive and sufficiently
strong only a minor part of the $^{41}$K impurity ends up
on the other side of the  $^{87}$Rb cloud. In Fig. \ref{f1} this effect
corresponds to a reflection coefficient $R$ that becomes different from zero.
Notably, when the inter-atomic interaction is strongly repulsive
the impurity of $^{41}$K is not able to enter the $^{87}$Rb cloud,
so that a barrier effect occurs (see movie \cite{movie-strongRep} for details).
The impurity then behaves as a classical object, similar
  to what observed with BEC solitons in presence of potential barriers
  much wider that their size \cite{marchant2013}.
In Fig. \ref{f1} this effect corresponds to a reflection coefficient $R$
close to one. However, if the initial displacement of the impurity
is increased so that to increase its initial potential energy,
the $^{41}$K cloud is able to penetrate the $^{87}$Rb cloud.
Due to the strong repulsive interaction a local minimum in the density
of $^{87}$Rb is observed in the correspondence of a sharp density peak
of the $^{41}$K impurity. The described situation is shown in the left panel
of Fig. \ref{f4}, where the normalized density profiles
$|\Psi_{Rb}(x,\tau)|^2/N_{Rb}$ and
 $|\Psi_K(x,\tau)|^2/N_K$  are reported for two different
values of $g_{12}/g_1$. The resulting moving hole in the  $^{87}$Rb BEC
is a dark soliton created by the interaction with the $^{41}$K impurity.
In this regime, the
  hole-impurity pair is similar to the dark-bright solitons observed in
  the superfluid counterflow of miscible condensates
  \cite{hamner2011}.
Notice that in Fig. \ref{f4} the upper panels are contour
plots of the normalized density profiles in the $(x,y)$ plane.
These contour plots are obtained adopting a Gaussian profile
along the $y$ axis with a width given by the characteristic length of
transverse harmonic confinement. However, for the sake of visibility,
the $y$ direction is plotted not in scale.
Movie \cite{movie-dark} displays the full dynamics of this dark soliton.

\subsubsection{Attractive inter-atomic strength.}

When the sign of $g_{12}$ is taken negative, the  \kq\
impurity enters and oscillates in the  \rb\ BEC; at each oscillation
a part of it is captured around the center,
so that after a certain time all the impurity gets confined around
this point. Increasing the magnitude of the negative $g_{12}$,
the number of oscillations prior to complete trapping decreases.
Moreover, when full confinement is reached, a local maximum in
the \rb\ density occurs, in the correspondence of the peak in the
\kq\ impurity. This moving peak is a bright soliton, shown in the
right panel of Fig. \ref{f4}. The full dynamics of this bright soliton
is reported in the movie, see \cite{movie-bright}.

\subsection{Capture and Localization effects}

\begin{figure}
\centering
\includegraphics[width=1\textwidth]{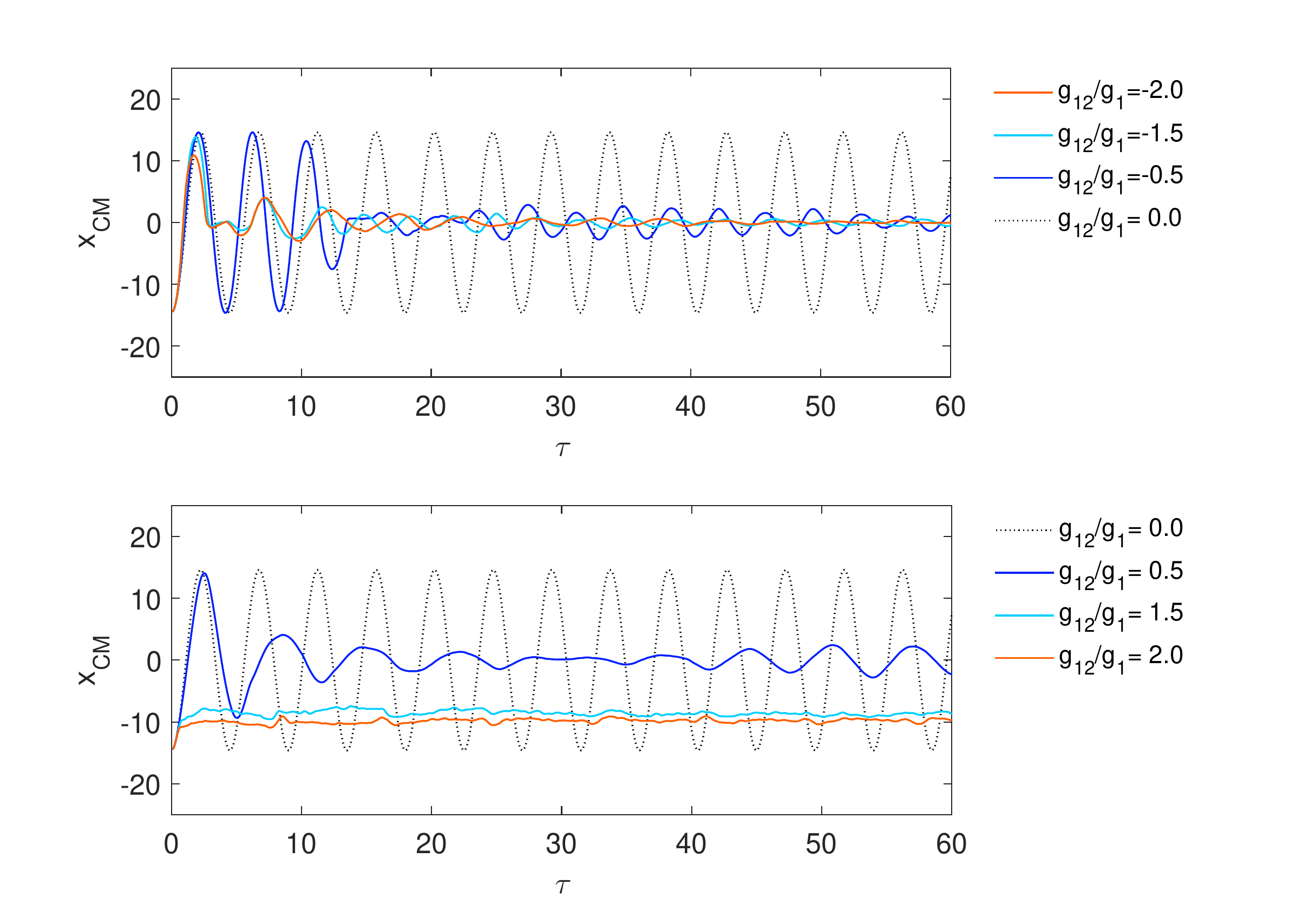}
\caption{Center of mass of the $^{41}$K cloud as function of time $\tau$ for attractive (upper panel) and repulsive (lower panel) values of $g_{12}/g_{1}$. $X_{CM}$ is plotted in unit of $a_{\parallel Rb}$.}
\label{fCM}
\end{figure}

Movies \gr{\cite{movie-0,movie-weakAttr,movie-weakRep,movie-strongRep,movie-dark, movie-bright}} show a broad range of phenomenology. Part of the rich behavior highlighted in the movies can be efficiently summarized in the dynamics of the  center of mass of the \kq\ cloud. In Fig. \ref{fCM} we plot
the center of mass of the $^{41}$K cloud as a function of time for several values of attractive (upper panel) and repulsive (lower panel) interaction $g_{12}/g_1$.  Fig. \ref{fCM} clearly shows how, for sufficiently strong repulsive values of $g_{12}/g_1$, the impurity cloud is completely blocked out (light-blue and orange line in the lower panel of Fig. \ref{fCM}), while in the opposite case, when the interaction is strongly attractive, the oscillations of $^{41}$K impurity is completely captured and confined at the center of the trap.

As pointed out in the previous section, for intermediate interaction strengths the system exhibit a more symmetric behavior. In both the attractive and the repulsive cases the impurity is always captured by the \rb\ BEC. This can be clearly seen in Fig. \ref{fCM} where for $g_{12}/g_1\in [-0.5, -2.0]$ (upper panel) and  $g_{12}/g_1=0.5$ (lower panel) the evolution of the center of mass is damped in amplitude and confined inside the \rb\ BEC (i.e. center of the trap). Interestingly, the ``capture mechanism'' seems to exhibit a set of common features among the different considered cases: (i) it occurs very quickly (about one oscillation cycle); (ii) smaller values of the interaction strength $|g_{12}|$ lead to a delay of the impurity capture; (iii) at least for intermediate interaction strengths, the impurity capture is always preceded by the appearance of an interference pattern in the density distribution of the impurity cloud. This sudden damping of the oscillation
has been verified to be
a rearrangement of internal energies, where part of the large initial kinetic energy of the impurity is handed over to the BEC cloud through inter-species interactions, hence producing a damping on the impurity oscillation. 

\begin{figure}
\centering
\includegraphics[width=1\textwidth]{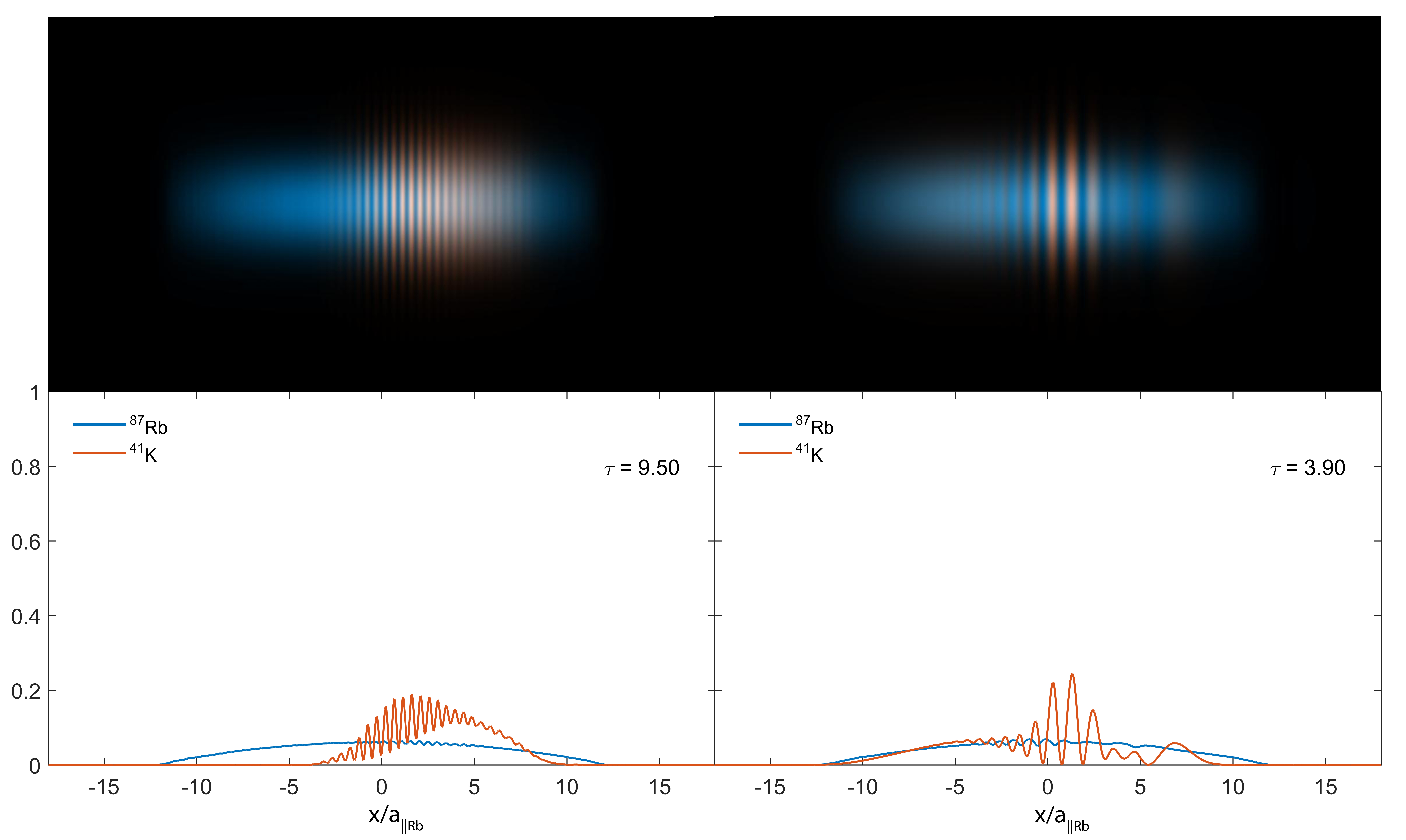}
\caption{Interference pattern in the $^{41}$K density distribution for attractive $g_{12}/g_{1}=-0.5$ (left panel) and repulsive $g_{12}/g_{1}=0.5$ (right panel). Upper panels give the contour plot of the normalized
density profiles in the $(x,y)$ plane. Lower panels give
the corresponding normalized axial density profiles
as a function of the adimensional axial coordinate $a_{\parallel Rb}$.
$\tau= \omega_{\parallel Rb} t$ is the adimensional time
with  $\omega_{\parallel Rb} = 389.6$ Hz and $a_{\parallel Rb}=1.37$ $\mu$m. Frames are taken from movies \cite{movie-weakAttr} and \cite{movie-weakRep}.}
\label{fint}
\end{figure}

In Fig. \ref{fint} we show two frames of the movies \cite{movie-weakAttr} and \cite{movie-weakRep}, for attractive $g_{12}/g_1=-0.5$ (left panel) and repulsive $g_{12}/g_1=0.5$ (right panel) cases in which an interference pattern appears. The interaction of the wave packet $\Psi_K(x,t)$ with the Rb condensate produces a reflected counter-propagating wave that, by quantum interfering with the incoming packet, produces the interference pattern shown in Fig \ref{fint}. This exclusively quantum phenomenon, previously encountered and studied
    in detail for instance in \cite{sheard2005,hamner2011}, proves to be driven by the interaction between the two atomic species $g_{12}$, and arises in both the attractive and repulsive cases leaving a direct signature of the energy transfer between the impurity and the condensate.
A difference in the spatial frequency of the interference pattern fringes is noticed between the attractive and repulsive cases with the former manifesting a generally higher spatial frequencies than the latter. However, the investigation of the true nature of such a difference goes beyond the goal of this paper and is 
left to future development.

A similar reflection-interference phenomenon appears at the boundary of the trap due to the reflection of the cloud between the parabolic walls. In that case, the intra-species interaction term $g_2$ provides that source of scattering between the \kq\ particles  capable of producing the reflected wave and the associated interference figure. We verified that if both $g_{2}=0$ and $g_{12}=0$ no interference patterns appear in the time evolution. In this case, the dynamical evolution of $|\Psi_K(x,t)|^2$ is perfectly described by the typical coherent-state picture.
%

The rich phenomenology described above reproduces the well-known
localization effects characterizing mixtures both in the absence
(see, e.g., \cite{PRA59})
and in the presence \cite{JPB49} of a superimposed optical lattice.
In the repulsive case, two species, fully mixed for
$g_{12}/ {\sqrt {g_1 g_2}} < \sigma \approx 1$ ($\sigma $ is determined
in \cite{PRA59} and \cite{JPB49}), separates when $g_{12}$ is sufficiently larger than
${\sqrt {g_1 g_2}}$ thus providing a spatial configuration where the density maximum of one species corresponds to density minimum of the other (the dark soliton in Fig. 4, lower left panel). A similar effect occurs if $g_{12}< 0$: when $|g_{12}|$  is sufficiently larger than ${\sqrt {g_1 g_2}}$, a configuration crops up in which the density maxima of both species perfectly overlap (local supermixing) due to the attractive interaction
(see Fig. 4, lower right panel).
Such configurations clearly emerge in the oscillations of the impurity in the Rb cloud. In particular, after the capture of the K impurity by the Rb cloud, one can observe how the final part of the oscillations shown in [33], [34] feature a stable bond of the K density maximum with the Rb dark soliton (bright soliton) in the presence of a strong repulsive (attractive) interaction.

\section{Conclusions}

In this paper we have analyzed the behavior of a system
characterized by quasi one-dimensional
Bose-Einstein condensate made of three hundreds \rb\ atoms interacting
with a bosonic impurity made of five \kq\ atoms, which starts from outside
the Bose condensate and collides on it. Despite the specific
physical system under investigation, the results we obtain are quite
general and, depending on the range and on the sign of the inter-species
interaction, different regimes can be identified.
Among them, the full reflection of the
impurity, the impurity trapping, and also the emergence of dark and bright
solitons in the Bose condensate. Preliminary experiments have measured
the reflection coefficient as a function of interatomic strength
in the system of Ref. \cite{catani2012} at finite temperature. Thus,
our zero-temperature theoretical predictions provide a useful benchmark for
forthcoming experimental and theoretical investigations
on impurity-BEC collisions. We think that our mean-field simulations,
based on coupled Gross-Pitaevskii equations, are quite reliable
but they could be improved adopting more sophisticated
time-dependent approaches where
quantum depletion is taken into account. In particular, when the \rb\ -- \kq\
scattering length is very large, the mean-field density-density interaction
is questionable and  beyond-mean-field effects could be relevant.
Finally, it is important to stress that a deeper connection between
impurity in bosonic atomic clouds and the solid-state polaron of Landau
and Fr\"olich can be obtained with a bosonic lattice
polaron \cite{grusdt}, i.e. a single-impurity atom confined
to a optical lattice and immersed in a homogeneous Bose-Einstein
condensate. We are now planning to investigate this difficult but
stimulating problem by using both mean-field and beyond-mean-field techniques.

\section*{Acknowledgements}
We thank Giacomo Lamporesi for useful discussions.
F.M. acknowledges funding from FP7 Cooperation STREP
Project EQuaM (Grant n. 323714). L.S. acknowledges Project BIRD164754 of
University of Padova for partial support.

\appendix
\section{Numerical computation}

Equations (\ref{GPE1}) and (\ref{GPE2}) are solved numerically by using a finite-difference predictor-corrector Crank-Nicolson algorithm \cite{salas1998}. Numerical discretization is performed on a fixed mesh-grid with constant spatial spacing $dx/a_{\parallel Rb}=1.50\times 10^{-3}$ and constant temporal spacing $dt/\omega_{\parallel Rb}=1.25\times 10^{-3}$. A single computation run, performed on a laptop
with an Intel i$7$ $2.90$ GHz processor and $16$ GB RAM, can take up to $40$ min with the above discretization.

To test the numerical accuracy of our results we checked the validity of the conservation laws for our solutions, namely, the conservation of energy $E$ and the conservation of total number of particles $N_{Rb}$ and $N_K$. Energy conservation is tested by computing the energy functional at any instant of time $t$
\begin{equation}
E(t)=\int \Psi_{Rb}^\star(x,t) \hat{H}_{Rb} \Psi_{Rb}(x,t)\,dx + \int \Psi_{K}^\star(x,t) \hat{H}_{K} \Psi_{K}(x,t)\,dx
\end{equation}
where $\hat{H}_{Rb}$ and $\hat{H}_{K}$ are the Hamiltonian operators on the right-hand side of equations (\ref{GPE1}) and (\ref{GPE2}) respectively.
Similarly, the test of the conservation of particle numbers is performed by checking the correct normalization of the wave-functions $\Psi_{Rb}(x,t)$ and $\Psi_{K}(x,t)$, verifying that equations (\ref{NR}) and (\ref{NK}) hold for any time $t$.
We verified that the normalization of the wave-functions is always almost perfectly conserved, and observed that,  with the given discretization also the energy is 
conserved ($>98\%$). Only for strong enough attractive interactions, the strong non-linearity of the soliton-solution leads to a decrease of energy conservation.
However, we verified that, by increasing time and space discretization, this problem is easily overcome at the expense of simulation-time, with no
qualitative change in our results.


\end{document}